\documentclass{iopart}
\usepackage{graphicx}
\usepackage{latexsym}
\usepackage{amssymb}

\jl{6}        
\eqnobysec    

\def\beq{\begin{equation}}
\def\eeq{\end{equation}}
\def\rmd{{\rm d}}

\def\pmb#1{\setbox0=\hbox{$#1$}%
  \kern-.025em\copy0\kern-\wd0
  \kern.05em\copy0\kern-\wd0
  \kern-.025em\raise.0433em\box0}

\def\bfbeta{\pmb{\beta}}

\def\dal{\mathop{\rlap{\hbox{$\sqcap$}}\sqcup}\nolimits}   

\begin{document}

\title[Gravitational induction]{Gravitational induction}

\author{
Donato Bini$^* {}^\S{}$,
Christian Cherubini${}^\S{}^\ddag$, Carmen Chicone$^\dag$ and Bahram Mashhoon${}^\diamond$
}
\address{
  ${}^*$\
Istituto per le Applicazioni del Calcolo ``M. Picone,'' CNR I-00161 Rome, Italy
}
\address{
  ${}^\S$\
  ICRA,
  University of Rome ``La Sapienza,'' I--00185 Rome, Italy
}

\address{
  $^\ddag$\
Nonlinear Physics and Mathematical Modeling Lab,  University
Campus Bio-Medico, I-00128 Rome, Italy }

\address{
  ${}^\dag$\
Department of Mathematics,
University of Missouri-Columbia, Columbia,
Missouri 65211, USA
}

\address{${}^\diamond$
Department of Physics and Astronomy,
University of Missouri-Columbia, Columbia,
Missouri 65211, USA}

\date{\today}

\begin{abstract}
   We study the linear post-Newtonian approximation to general relativity
known as gravitoelectromagnetism (GEM); in particular, we examine the
similarities and differences between GEM and electrodynamics.
Notwithstanding some significant differences between them, we find that a special nonstationary metric in
GEM can be employed to show {\it explicitly} that it is possible to introduce
gravitational induction within GEM in close analogy with Faraday's law of
induction and Lenz's law in electrodynamics. Some of the physical
implications of gravitational induction are briefly discussed.
\end{abstract}

\pacs{04.20.Cv}

\section{Introduction}
In electromagnetism, the combined dynamics of charged particles and electromagnetic field are
consistently described by Maxwell's field equations and the Lorentz force law.
However, in the linear perturbation approach to gravitoelectromagnetism (GEM), one recovers the Maxwell equations for the GEM field, but the corresponding Lorentz force is recovered, to first order in ${\mathbf v}/c$, only when we deal with a {\it stationary} GEM field.
This explains why some authors (see, for instance, \cite{ciufowhee,harris}) have treated GEM only for stationary fields and the issue of existence of gravitational induction in analogy with Faraday's law of induction is therefore absent in such treatments; moreover, it has been argued recently that in general relativity such an analogy does not even exist \cite{costaetal}. On the other hand, time-varying GEM fields have been {\it implicitly} considered by many authors (see, for instance, \cite{bracavtho}--\cite{mlrtart2}). In fact, some gravitational Faraday experiments were proposed in \cite{bracavtho} based on the existence of gravitational induction in analogy with electrodynamics.
The purpose of the present paper is to show {\it explicitly} that general relativity does indeed contain induction effects; these turn out to be, despite the differences that have been mentioned, on the whole closely analogous to electromagnetic induction effects.

In general, GEM covers those aspects of general relativity that can be best explained via an electromagnetic analogy.
In this paper, we work mainly within the linear GEM scheme; therefore, it is necessary to review briefly the relevant aspects of this linear post-Newtonian approximation to general relativity \cite{mashh1} in order to render the present paper essentially self-contained. Consider the curved spacetime generated by a localized slowly rotating \lq\lq nonrelativistic" astronomical source. In the linear approximation, the spacetime metric can be written as $g_{\mu\nu}=\eta_{\mu\nu}+h_{\mu\nu}$, where $\eta_{\mu\nu}$ is the Minkowski metric tensor with signature $+2$ in our convention  and $h_{\mu\nu}$ is a first-order perturbation.
Under a slight transformation of the background coordinates $x^\mu=(ct,{\mathbf x})$, $x^\mu \mapsto x^\mu-\epsilon^\mu$, the gravitational potentials $h_{\mu\nu}$ transform as $h_{\mu\nu}\mapsto h_{\mu\nu}+\epsilon_{\mu, \nu}+\epsilon_{\nu ,\mu}$.
Henceforth, the potentials are considered to be gauge dependent, while the background global inertial coordinate system is in effect fixed. The spacetime curvature is, however, gauge invariant. It is useful to introduce the trace-reversed potentials $\bar h_{\mu\nu}=h_{\mu\nu}-\frac12 h \eta_{\mu\nu}$ with $h={\rm tr}(h_{\mu\nu})$.
Imposing the transverse gauge condition $\bar h^{\mu\nu}{}_{,\nu}=0$, the gravitational field equations take the form
\beq
\label{eq:1}
\dal \bar h_{\mu\nu}=-\frac{16\pi G}{c^4}T_{\mu\nu}\ .
\eeq
The general solution of (\ref{eq:1}) is given by the special retarded solution
\beq
\label{eq:2}
 {\bar h}_{\mu\nu}=\frac{4G}{c^4}\int \frac{T_{\mu\nu}(ct-|{\mathbf x}-{\mathbf x}'|, {\mathbf x}')}{|{\mathbf x}-{\mathbf x}'|}\rmd^3 x'\ ,
\eeq
plus a general solution of the homogeneous wave equation that we simply ignore in this work. In the linear GEM approach, all terms of $O(c^{-4})$ are neglected in the metric tensor. It then follows from Eq. (\ref{eq:2}) that for the sources under consideration here
$\bar h_{00}=4\Phi/c^2$,
$\bar h_{0i}=-2A_i/c^2$ and $\bar h_{ij}=O(c^{-4})$, where $\Phi(t,{\mathbf x})$ is the gravitoelectric potential and ${\mathbf A}(t,{\mathbf x})$ is the gravitomagnetic vector potential. The spacetime metric is thus given by
\beq
\label{eq:3}
\fl\qquad
\rmd s^2= -c^2 \left(1-2\frac{\Phi}{c^2}\right)\rmd t^2 -\frac4c ({\mathbf A}\cdot \rmd {\mathbf x})\rmd t +
 \left(1+2\frac{\Phi}{c^2}\right)\delta_{ij}\rmd x^i \rmd x^j\ ,
\eeq
where far from the source the dominant contributions to the GEM potentials can be expressed as
\beq
\label{eq:4}
\Phi = \frac{GM}{r}, \qquad {\mathbf A}=\frac{G}{c}\frac{{\mathbf J}\times {\mathbf x}}{r^3}\ .
\eeq
Here $M$ and ${\mathbf J}$ are the inertial mass and angular momentum of the source, $r=|{\mathbf x}|$,
$r\gg GM/c^2$ and $r\gg J/(Mc)$. Let us note that the gauge condition implies that
\beq
\label{eq:5}
\frac{1}{c}\partial_t \Phi+\nabla \cdot \left( \frac12 {\mathbf A}\right)=0.
\eeq
This is related to the conservation of mass-energy of the source via Eq. (\ref{eq:1}). That is, let
$T^{00}=\rho c^2$ and $T^{0i}=cj^i$, where $j^\mu=(c\rho,{\mathbf j})$ is the mass-energy current of the source; then, Eq. (\ref{eq:5}) is equivalent to $j^\mu{}_{,\mu}=0$.
It is possible to define the gravitoelectric field ${\mathbf E}$ and the gravitomagnetic field ${\mathbf B}$
in close analogy with electrodynamics
\beq
\label{eq:6}
{\mathbf E}=-\nabla \Phi-\frac{1}{c}\partial_t \left( \frac12 {\mathbf A}\right), \qquad {\mathbf B}=\nabla \times {\mathbf A}\ .
\eeq
It follows from these definitions that
\beq
\label{eq:7}
\nabla \times {\mathbf E}=-\frac{1}{c}\partial_t \left( \frac12 {\mathbf B}\right),\qquad
\nabla \cdot \left(\frac12 {\mathbf B} \right)=0,
\eeq
while the gravitational field equations (\ref{eq:1}) imply
\beq
\label{eq:8}
\nabla \cdot {\mathbf E}=4\pi G \rho, \qquad \nabla \times \left( \frac12 {\mathbf B}\right)=
\frac{1}{c}\partial_t {\mathbf E}+\frac{4\pi G}{c}{\mathbf j}.
\eeq
These are the Maxwell equations for the GEM field.
The particular form of these equations is based on a special convention \cite{mas93} that makes it possible to employ the standard results of classical electrodynamics in the GEM framework. This is accomplished by assuming that the source has gravitoelectric charge $Q_E=GM$ and gravitomagnetic charge $Q_B=2GM$.
Moreover, a test particle of inertial mass $m$ has gravitoelectric charge $q_E=-m$ and gravitomagnetic charge $q_B=-2m$ in this convention. The signs of $(q_E,q_B)$ are opposite to those of $(Q_E,Q_B)$ due to the attractive nature of gravity; furthermore, the ratio of gravitomagnetic charge to the gravitoelectric charge is always 2, as the linear approximation of general relativity involves a spin-2 field. This circumstance is consistent with the fact that the ratio of the magnetic charge to the electric charge of a particle is unity in Maxwell's spin-1 electrodynamics. We note that the magnetic charge employed here is different from the magnetic monopole strength, which is always strictly zero throughout this work.

Given Maxwell's equations for the electromagnetic field, Faraday's law of induction simply follows,
for instance, from the consideration of the temporal variation of the magnetic flux linking a static closed circuit.
A similar approach in the GEM case would fail, however, as the line integral of ${\mathbf E}$ along the closed circuit
does not in general correspond to the work done by the gravitational field of the source.
This is the crucial point and to clarify the situation, it is therefore necessary to investigate
the motion of a free test particle in the linear GEM scheme.

We must now discuss the analogue of the Lorentz force law in our linear GEM framework. The geodesic equation for the motion of a free test particle is
\beq
\label{eq:9}
\frac{\rmd u^\mu}{\rmd \tau} +\Gamma^\mu{}_{\rho\sigma}u^\rho u^\sigma=0\ ,
\eeq
where $\tau/c$ is the proper time and $u^\mu=\rmd x^\mu/\rmd \tau$ is the unit four-velocity vector of the test particle.
The Christoffel symbols are given by
\begin{eqnarray}
\label{eq:10}
c^2\Gamma^0{}_{0\mu}&=&-\Phi_{,\mu},\qquad\qquad\,\,\,\, c^2\Gamma^0{}_{ij}=2A_{(i,j)}+\delta_{ij}\Phi_{,0}\ ,  \\
\label{eq:11}
c^2\Gamma^i{}_{00}&=&-\Phi_{,i}-2A_{i,0} ,\qquad c^2\Gamma^i{}_{0j}=\delta_{ij}\Phi_{,0}+\epsilon_{ijk}B^k\ , \\
\label{eq:12}
c^2\Gamma^i{}_{jk}&=&\delta_{ij}\Phi_{,k}+\delta_{ik}\Phi_{,j} -\delta_{jk}\Phi_{,i}\ .
\end{eqnarray}
The geodesic equation can be reduced via $u^\mu=\gamma (1,\bfbeta )$ with $\bfbeta={\mathbf v}/c$ to
\begin{eqnarray}
\label{eq:13}
\frac{c}{\gamma}\frac{\rmd \gamma}{\rmd t}&=&(1-\beta^2)\Phi_{,0}+2\beta^i[\Phi_{,i}-A_{(i,j)}\beta^j],  \\
\frac{\rmd v^i}{\rmd t}&=&(1+\beta^2)\Phi_{,i}-2({\pmb \beta}\times {\mathbf B})_i+2A_{i,0}\nonumber \\
\label{eq:14}
&& -\beta^i (3-\beta^2)\Phi_{,0}+2\beta^i\beta^j [A_{(j,k)}\beta^k -2\Phi_{,j}]\ .
\end{eqnarray}
Moreover, $u^\mu u_\mu=-1$ implies that
\beq
\label{eq:15}
\frac{1}{\gamma^2}=1-\beta^2-\frac{2}{c^2}(1+\beta^2)\Phi+\frac{4}{c^2}{\pmb\beta}\cdot {\mathbf A}.
\eeq
For a stationary source ($\partial_t \Phi=0$ and
$\partial_t {\mathbf A}=0$), Eq. (\ref{eq:14}) reduces to
\beq
\label{eq:16}
m\frac{\rmd {\mathbf v}}{\rmd t}=-m{\mathbf E}-2m \frac{{\mathbf v}}{c}\times {\mathbf B},
\eeq
when velocity-dependent terms of order higher that $\beta=v/c$ are neglected.
In the case of a general nonstationary
source, however, the equation of motion (\ref{eq:14}) does not correspond to the Lorentz force law and
this implies that the electromotive force does not in general have a simple analogue in GEM.

Though the gravitational analogue of the Lorentz force law has a more
complicated form in GEM, we intend to show via a special nonstationary GEM metric that
induction effects can still
exist in close analogy with electrodynamics. The motivation for our approach
comes from a detailed consideration of the gravitomagnetic clock effect.
This is briefly discussed in the next section.

\section{A nonstationary GEM metric}

We start our analysis with a brief discussion of the gravitomagnetic clock effect, since there is an important heuristic connection between gravitational induction and this effect. Consider circular equatorial geodesics about a Kerr source of mass $M$ and angular momentum $J$.
The exterior spacetime metric is given by
\begin{eqnarray}
\label{eq:17}
\rmd s^2&=& -c^2 \rmd t^2 +\frac{\Sigma}{\chi}(\rmd \rho^2 +\chi \rmd \theta^2)+(\rho^2+a^2)\sin^2\theta \rmd \phi^2 \nonumber \\
&& +\frac{2\hat M\rho}{\Sigma}(c\rmd t-a\sin^2 \theta \rmd \phi)^2,
\end{eqnarray}
where $\hat M=GM/c^2$, $a=J/(Mc)>0$ is the specific angular momentum of the Kerr source and
\beq
\label{eq:18}
\Sigma=\rho^2+a^2\cos^2\theta, \qquad \chi=\rho^2-2\hat M \rho +a^2\ ,
\eeq
in Boyer-Lindquist coordinates. The geodesic equation for a circular equatorial orbit reduces to
\beq
\label{eq:19}
\frac{\rmd t}{\rmd \phi}=\pm \frac{1}{\omega_K}+\frac{a}{c}\ ,
\eeq
where $\omega_K$ is the Keplerian frequency, $\omega_K=(GM/\rho^3)^{1/2}$, for the orbit with fixed \lq\lq radius" $\rho>2\hat M$ and $\theta=\pi/2$. The upper (lower) sign in Eq. (\ref{eq:19}) refers to a co-rotating (counter-rotating)
orbit with respect to the sense of rotation of the Kerr source. It follows from (\ref{eq:19}) that
\beq
\label{eq:20}
t_\pm=\frac{2\pi}{\omega_K}\pm 2\pi \frac{a}{c}\ ,
\eeq
where $t_+$ ($t_-$) is the period of prograde (retrograde) circular motion in terms of the proper time of the static inertial observers that are infinitely far from the source. We are interested in physical situations where the orbital motion is far from the source, i.e. $\rho \gg 2 \hat M$ and $\rho \gg J/(cM)$, so that geodesic motion is possible in opposite directions for the same orbital \lq\lq radius" $\rho$. Then $t_+-t_-=4\pi J/(Mc^2)$ illustrates the gravitomagnetic clock effect.
This remarkable result, which is independent of $G$ and $\rho$, holds to lowest order for the proper times of clocks in orbit around the source as well. It has already been discussed in a number of papers (see, for example, \cite{mashh2,co-mas,bijamas}); therefore, we concentrate here on the fact that the prograde motion is {\it slower} than the retrograde motion. Specifically, let $v_\pm=2\pi \rho/t_\pm$ be the relevant speed of motion according to the static inertial observers at spatial infinity ($\rho\to \infty$); then, to first order in $a\omega_K/c\ll 1$,
\beq
\label{eq:21}
v_\pm\approx v_K\mp \frac{GJ}{c^2\rho^2}\ ,
\eeq
where $v_K=\rho\omega_K$.

Imagine an ensemble of identical Kerr spacetimes except for different magnitudes of $J$. As $J$ increases in this ensemble, $v_+$ decreases and  $v_-$ increases. Let us first note that this circumstance cannot be interpreted as \lq\lq inertial induction" \cite{einst}, as the effect is simply the opposite of what such a Machian interpretation would predict --- we return to this subject in section 5. On the other hand, if one could turn this kinematic situation into a dynamic one in terms of the temporal variation of $J$, then one could at least heuristically interpret the change in speeds in terms of induced currents due to a time-varying flux of the gravitomagnetic field. To this end, we need a solution of the field equations of general relativity that would correspond to a Kerr solution but with a time-varying $J$. It turns out that such a solution exists \cite{mastobepub}, but only in a rather approximate form within the linear GEM framework. We therefore turn to a description of this solution (see also Appendix A).

Let us first note that the Kerr solution can be put into the form of metric (\ref{eq:3}) with potentials (\ref{eq:4}) once the isotropic radial coordinate $r$
\beq
\label{eq:22}
\rho=r\left(1+\frac{\hat M}{2r}\right)^2
\eeq
is introduced in Eq. (\ref{eq:17}) and the resulting metric is linearized in $\hat M/r$ and $a/r$ with $x=r\sin\theta \cos \phi$, $y=r\sin\theta\sin\phi$ and $z=r\cos \theta$.

Consider next a spacetime metric of the GEM form (\ref{eq:3}) with potentials
\beq
\label{eq:23}
\Phi=\frac{GM}{r}, \qquad {\mathbf A}=\frac{G}{c}(J_0+J_1t)\frac{\hat {\mathbf J}\times {\mathbf x}}{r^3}\ ,
\eeq
where the magnitude of the proper angular momentum  of the source varies linearly with time, i.e.
${\mathbf J}(t)=(J_0+J_1t)\hat {\mathbf J}$. Henceforth we assume that $\hat {\mathbf J}=\hat {\mathbf z}$ and $J_0\ge 0$. Substituting potentials (\ref{eq:23}) in the GEM field equation (\ref{eq:1}) and gauge condition (\ref{eq:5}), we find that the latter is satisfied and the former gives the {\it effective} source of the spacetime metric in the form
\beq
\label{eq:24}
T_{00}=Mc^2\delta({\mathbf x}), \qquad
T_{0i}=\frac12 c [{\mathbf J}(t)\times \nabla ]_i \delta({\mathbf x}) \ .
\eeq
It follows from the dynamical equations for the source
\beq
\label{eq:25}
T^{\mu\nu}{}_{,\nu}=0
\eeq
that
\beq
\label{eq:26}
T_{ij}=-\frac{1}{8\pi}\left[(\dot {\mathbf J}\times \nabla)_i \nabla_j \frac1r +
(\dot {\mathbf J}\times \nabla)_j \nabla_i \frac1r
\right]\ ,
\eeq
where
$\dot {\mathbf J}=\rmd {\mathbf J}/ \rmd t=J_1 \hat {\mathbf J}$ is a constant vector. While the mass-energy current (\ref{eq:24}) is confined to the origin of spatial coordinates, the stresses (\ref{eq:26}) are distributed throughout space and fall off as $r^{-3}$ for $r\to\infty$; however, this unusual circumstance has no impact on the viability of this nonstationary GEM spacetime, since $T_{ij}$ is independent of time and Eq. (\ref{eq:2}) then implies that $\bar h_{ij}=O(c^{-4})$ for the stresses (\ref{eq:26}). Thus this special source
satisfies the requirements of the linear GEM scheme and hence generates an acceptable GEM field.
That is, metric (\ref{eq:3}) together with potentials (\ref{eq:23}) represents
a solution of the linearized field equations that is valid at the first post-Newtonian order of approximation.
Further details about this time-varying but nonradiative solution are given in \cite{mastobepub} and Appendix A.
It is clear that the linear perturbation approach eventually breaks down over a sufficiently extended period of time due to the linear temporal variation of the gravitomagnetic potential in Eq. (\ref{eq:23}). On the other hand, we wish to avoid any complications associated with the instants of time at which the temporal variation is switched on and off. We therefore consider a certain time interval after the temporal variation is switched on and before it is switched off such that our linear GEM approach is valid; in fact, we will always work within this interval of time for which $2|{\mathbf  A}|\ll c^2$.

There exist radiative solutions of Einstein's field equations of the type originally due to Vaidya in which the mass (and hence possibly angular momentum) of the source can vary with time due to the emission or absorption of radiation. Such solutions are not of interest here. Instead, we concentrate on nonradiative solutions in which the angular momentum slowly varies with time. For instance, the Earth slowly loses angular momentum with time due mainly to tidal friction. Conservation of angular momentum implies that the orbital angular momentum of the Moon about the Earth increases in this process. The Earth-Moon distance thus slowly increases; the rate of orbital expansion at present is about $4$ cm per year. The special nonstationary solution can be employed to discuss the physical implications of the temporal variability of the gravitomagnetic field. In the case of the GP-B experiment performed in orbit about the Earth, for example, the influence of such variability turns out to be negligible \cite{mastobepub}. However, our considerations may be of interest in the astrophysics of rotating gravitationally collapsed
configurations that exhibit variability.

The next section is devoted to the illustration of gravitational
induction and Lenz's law using the special nonstationary GEM solution (\ref{eq:23}).

\section{GEM induction and Lenz's law}

The special nonstationary spacetime given by potentials (\ref{eq:23}) involves a static gravitoelectric field and a linearly time-varying gravitomagnetic field
\begin{eqnarray}
\label{eq:27}
{\mathbf E}&=& \frac{GM{\mathbf x}}{r^3}-\frac{G}{2c^2}\frac{\dot {\mathbf J}\times {\mathbf x}}{r^3},\\
\label{eq:28}
{\mathbf B}&=& \frac{G}{c}(J_0+J_1t)\frac{1}{r^3}[3\,(\hat {\mathbf J}\cdot \hat {\mathbf x})\hat {\mathbf x}-\hat {\mathbf J}]\ .
\end{eqnarray}
Thus the gravitational analogue of the displacement current is zero in this case. To first order in ${\mathbf v}/c$, the analogue of the Lorentz force law is given by
\beq
\label{eq:29}
m\frac{\rmd {\mathbf v}}{\rmd t}=-m\, {\pmb {\mathcal E}}-\frac{2m}{c}{\mathbf v}\times {\mathbf B},
\eeq
where ${\pmb {\mathcal E}}$ can be expressed as
\begin{eqnarray}
\label{eq:30}
{\pmb {\mathcal E}}&=& \frac{GM{\mathbf x}}{r^3}
-\frac{2G}{c^2}\frac{\dot {\mathbf J}\times {\mathbf x}}{r^3} \ .
\end{eqnarray}
The distinction between ${\mathbf E}$ and ${\pmb {\mathcal E}}$ in general relativity is the root of the difference between the electromagnetic and gravitational inductions. A free particle initially at rest picks up an azimuthal speed due to the force term $(2m/c)\partial_t {\mathbf A}$ in Eq. (\ref{eq:29}). The general motion of a free test particle in this nonstationary spacetime is studied in some detail in the next section; moreover, certain aspects of the motion of spinning particles and light rays have been briefly considered in \cite{mastobepub}.

Consider now a closed circuit in the equatorial plane of the variable rotating source as depicted in Figure 1.
\begin{figure}
\begin{center}
\includegraphics[scale=0.35]{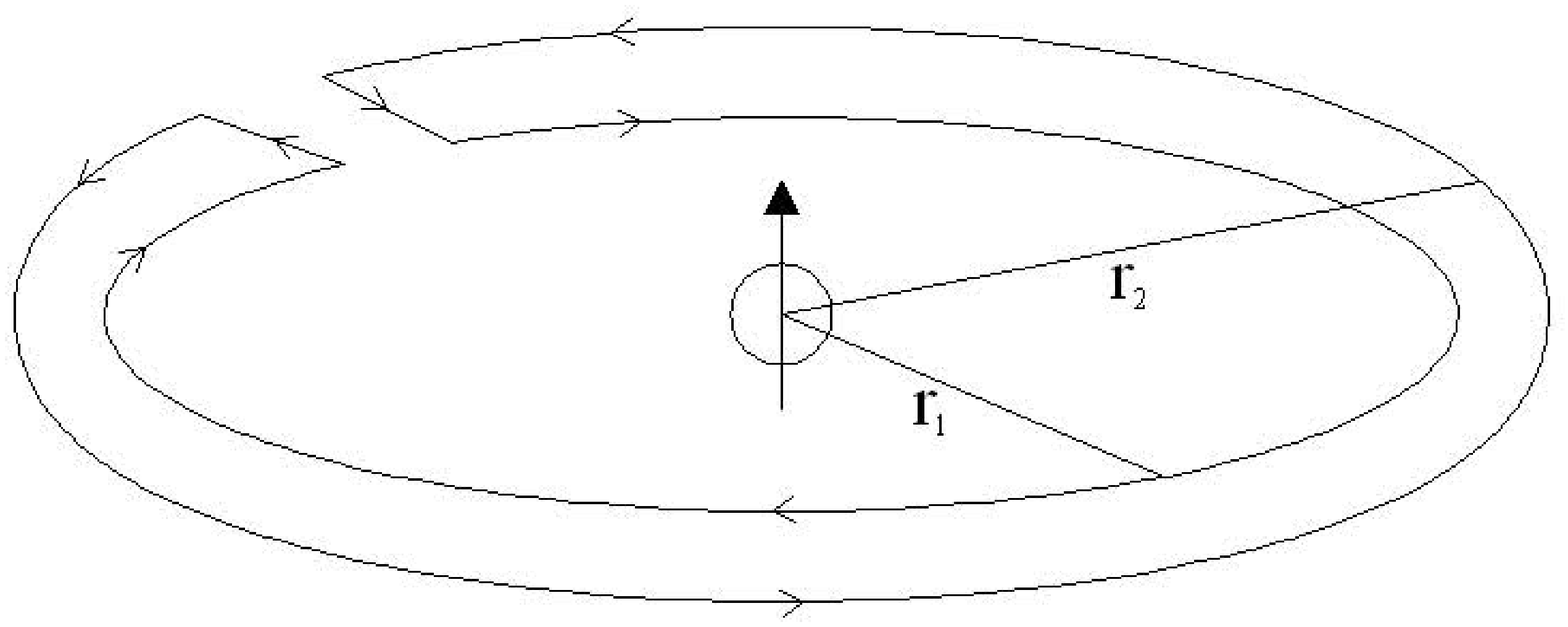}
\caption{Schematic diagram of the closed equatorial annular loop for a thought experiment that illustrates gravitational induction. The loop is assumed to be sufficiently far from a nonstationary rotating source.}
\end{center}
\end{figure}
The circuit is bounded by an inner circle of radius $r_1$ and an outer circle of radius $r_2$. The radial parts of the circuit are infinitesimally close to each other. For the sake of concreteness, let the circuit consist of a perfect fluid that is at rest and fills a very thin tube. Once the circuit is placed in the time-varying gravitomagnetic field as in Figure 1, the flux of this field through the circuit is
\beq
\label{eq:31}
{\mathcal F}=\int {\mathbf B}\cdot \rmd {\mathbf S}=-\frac{2\pi G}{c}J(t) \left(\frac{1}{r_1}-\frac{1}{r_2}\right)\ .
\eeq
Moreover, we note that
\beq
\label{eq:32}
\oint {\mathbf E}\cdot \rmd {\pmb \ell}=\frac{\pi G \dot J}{c^2}\left(\frac{1}{r_1}-\frac{1}{r_2}\right)\ ,
\eeq
where $\rmd {\pmb \ell}$ is an element of the circuit in the direction depicted in Figure 1. Thus it follows from Eqs.
(\ref{eq:31}) and (\ref{eq:32}) that
\beq
\label{eq:33}
\oint {\mathbf E}\cdot \rmd {\pmb \ell}=-\frac{1}{2c}\frac{\rmd {\mathcal F}}{\rmd t}\ ,
\eeq
which, as expected, is in accordance with Maxwell's equations for the GEM field. It turns out, however, that Eq. (\ref{eq:33}) is not the analogue of Faraday's law of induction in the gravitational case.

In electrodynamics, the quantity evaluated in Eq. (\ref{eq:32}) would be the electromotive force (emf), which is the amount of work done per unit electric charge. The gravitational analogue of this concept would be the amount of work done per unit gravitoelectric charge. We therefore define --- on the basis of Eq. (\ref{eq:29}) --- the {\it gravitomotive force } (gmf) to be
\beq
\label{eq:34}
{\mathcal G}=\oint {\pmb {\mathcal E}}\cdot \rmd {\pmb \ell}\ ,
\eeq
where ${\pmb {\mathcal E}}$ is given by Eq. (\ref{eq:30}), since the ${\mathbf v} \times {\mathbf B}$ term does not contribute to the work. Thus the gravitomotive force is the circulation of ${\pmb {\mathcal E}}$ rather than ${\mathbf E}$. Calculating the gmf for the loop in Figure 1, we find
\beq
\label{eq:35}
{\mathcal G}=\frac{4\pi G\dot J}{c^2}\left(\frac{1}{r_1}-\frac{1}{r_2}\right)\ .
\eeq
Thus the gravitational analogue of Faraday's law of induction turns out to be
\beq
\label{eq:36}
{\mathcal G}=-\frac{2}{c}\frac{\rmd {\mathcal F}}{\rmd t}\ .
\eeq
Indeed, the actual induced current is four time larger than that given by Eq. (\ref{eq:33}), which is evident from
the factor of four difference between the coefficients of $\partial_t {\mathbf A}$ in Eqs. (\ref{eq:30}) and (\ref{eq:27}).
Nevertheless, these considerations show theoretically that gravitational induction exists and --- within the linear GEM framework --- is closely analogous to the Faraday law of induction in electrodynamics.

It remains to show that the direction of the induced current is such as to oppose the change that caused it.
Let us first remark that for a long straight line supporting a mass current $I$, Eq. (\ref{eq:8}) implies that the gravitomagnetic field has closed circular field lines of radius $r$ around the line current as in electrodynamics; moreover, the magnitude of the field is $4GI/(cr)$ and its direction follows from the usual right-hand rule. In fact, this is the gravitational analogue of the Biot-Savart law of electrodynamics. For a moving test particle of mass $m$, the gravitoelectric charge is $-m$; hence, the direction of its current is {\it opposite} to its velocity. As the gravitomagnetic field increases ($\dot J>0$), the fluid (in the tube in Figure 1) begins to co-rotate from rest due to the induced azimuthal acceleration $(2/c)\partial_t {\mathbf A}$. The speed of this co-rotation is larger in the inner circle than in the outer circle of the loop, so that a clockwise motion develops in the loop corresponding to an induced counter-clockwise current and hence positive flux through the loop that opposes the increasingly negative flux of the gravitomagnetic field of the source. This is the gravitational analogue of Lenz's law of electrodynamics.

Our treatment of GEM induction and Lenz's law has been based upon our specific thought experiment and the use of the special nonstationary solution. The main results are, however, quite general.
Assuming $\partial_t \Phi=0$ and working to first order in ${\mathbf v}/c$, we find that the equation of motion (\ref{eq:29}) is generally valid with
${\pmb {\mathcal E}}=-\nabla\Phi-(2/c)\partial_t {\mathbf A}$ and ${\mathbf B}=\nabla \times {\mathbf A}$ for a nonstationary ${\mathbf A}(t,{\mathbf x})$. Thus with the definitions of the gravitomotive force ${\mathcal G}$ and the gravitomagnetic flux ${\mathcal F}$, Eq. (\ref{eq:36}) is generally satisfied. This is then the general form of the law of gravitational induction in GEM and the minus sign on the right-hand side of Eq. (\ref{eq:36}) is in accordance with Lenz's law.

As demonstrated in section 2, in an {\it ensemble} of stationary Kerr spacetimes with varying $J$,
the speeds of prograde (retrograde) circular equatorial orbits decrease (increase) with increasing $J$.
It is interesting to demonstrate this effect {\it dynamically} using the nonstationary linearized Kerr spacetime.
This is the purpose of the next section.

\section{Motion in a time-varying gravitomagnetic field}

We start with the equation of motion (\ref{eq:14})
adapted to potentials (\ref{eq:23}), namely,
\begin{eqnarray}
\label{eq:37new}
\frac{\rmd {\mathbf v}}{\rmd t}+\frac{GM{\mathbf x}}{r^3}&=& \frac{GM}{c^2r^3}[4({\mathbf x}\cdot {\mathbf v}){\mathbf v}-v^2{\mathbf x}]+\frac{2G}{c^2}\frac{\dot {\mathbf J}\times {\mathbf x}}{r^3}-\frac{2}{c}{\mathbf v}\times {\mathbf B}\nonumber \\
&& -\frac{6GJ(t)}{c^4r^5}[\hat {\mathbf J}\cdot ({\mathbf x}\times {\mathbf v})]({\mathbf x}\cdot {\mathbf v}){\mathbf v}
\end{eqnarray}
and inquire whether {\it equatorial circular orbits} are possible in this case. With  $x=r_0\cos\phi$, $y=r_0\sin \phi$ and $z=0$, Eq. (\ref{eq:37new}) reduces to
\begin{eqnarray}
\label{eq:37}
\ddot \phi&=& \frac{2GJ_1}{c^2r_0^3}\ , \\
\label{eq:38}
v^2&=& \frac{GM}{r_0}\left(1+\frac{v^2}{c^2}\right)- \frac{2GJ(t)}{c^2r_0^2}v\ ,
\end{eqnarray}
where $v=r_0\dot \phi$ and $\dot \phi=\rmd \phi/ \rmd t$.
Solving the quadratic equation for $v$ to linear order in Newton's constant, differentiating the outcome with respect to time and comparing the result with Eq. (\ref{eq:37}), we find
that Eqs. (\ref{eq:37})--(\ref{eq:38}) are inconsistent so long as $J_1\not =0$. Thus no circular geodesic orbits exist in the equatorial plane of the source for $J_1\not =0$. On the other hand, for $J_1=0$, we find from Eqs. (\ref{eq:37})--(\ref{eq:38}) that for a circular orbit
\beq
\label{eq:39}
v\approx \pm \sqrt{\frac{GM}{r_0}}- \frac{GJ_0}{c^2r_0^2}
\eeq
in agreement with Eq. (\ref{eq:21}).
Here the upper (lower) sign refers to a prograde (retrograde) orbit and terms beyond the linear order in Newton's gravitational constant have been neglected. Once $J$ starts to increase linearly with time, the prograde (retrograde) equatorial circular orbits tend to spiral outward (inward); moreover, the azimuthal speeds of prograde (retrograde) orbits decrease (increase) according to Eq. (\ref{eq:37new}), as would be expected from our discussion of the gravitomagnetic clock effect in section 2.
These results follow from the perturbative studies of Eq. (\ref{eq:37new}) that are presented in the rest of this section.

It is useful to write Eq. (\ref{eq:37new}) as
\beq
\label{eq:41}
\frac{\rmd {\mathbf v}}{\rmd t}+\frac{GM{\mathbf x}}{r^3}={\mathbf F}\ ,
\eeq
where ${\mathbf F}$ is the linear GEM relativistic perturbing acceleration.
Let us digress here and note that to $O(c^{-2})$, ${\mathbf F}$ should also contain the nonlinear post-Newtonian term $4G^2M^2{\mathbf x}/(c^2r^4)$, which we have neglected in our linear treatment. For  ${\mathbf F}=0$, the test particle follows a Keplerian orbit; in this case, the Newtonian energy $E_N$, orbital angular momentum ${\mathbf L}_N$ and the Runge-Lenz vector ${\mathbf R}_N$ (all per unit mass of the test particle) are conserved. These are given by
\begin{eqnarray}
\label{eq:42}
E_N&=& \frac12 v^2-\frac{GM}{r}\ , \quad {\mathbf L}_N={\mathbf x}\times {\mathbf v}, \\
\label{eq:43}
{\mathbf R}_N&=& {\mathbf v}\times {\mathbf L}_N -\frac{GM}{r}{\mathbf x}\ ,
\end{eqnarray}
so that in the presence of the perturbation ${\mathbf F}$, we have
\begin{eqnarray}
\label{eq:44}
\frac{\rmd E_N}{\rmd t}&=& {\mathbf F}\cdot {\mathbf v}\ , \quad \frac{\rmd {\mathbf L}_N}{\rmd t} ={\mathbf x}\times {\mathbf F}, \\
\label{eq:45}
\frac{\rmd {\mathbf R}_N}{\rmd t}&=& {\mathbf F}\times ({\mathbf x}\times {\mathbf v})+ {\mathbf v}\times ({\mathbf x}\times {\mathbf F})\ .
\end{eqnarray}

For our present purpose, it suffices to concentrate on the energy equation and compute ${\mathbf F}\cdot {\mathbf v}$ using Eq. (\ref{eq:37new}) along any unperturbed Keplerian circular orbit around the source. The unperturbed orbital plane can have an arbitrary inclination angle $i$ as in Figure 2. For a circular orbit ${\mathbf x}\cdot {\mathbf v}=0$ and hence the Newtonian energy equation reduces to
\beq
\label{eq:46}
\frac{\rmd }{\rmd t}\left( \frac12 v^2-\frac{GM}{r}\right)=\frac{2GL_N}{c^2r^3}\dot J \cos i\ ,
\eeq
where $L_N=|{\mathbf L}_N|$. Thus for $\dot J \cos i>0$, the Newtonian energy increases and hence the circular orbit tends to spiral outward, but for
$\dot J \cos i<0$, the orbit spirals inward; moreover, for a polar orbit ($\cos i =0$), it remains unchanged within the orbital plane regardless of $\dot J$.

More generally, let us assume that the unperturbed orbit is an ellipse with semimajor axis $R_0$ and eccentricity $e$ in the $(X,Y)$ plane. The correspondence between
the $(x, y, z)$ and $(X, Y, Z)$ coordinate systems is illustrated in Figure 2.
The ellipse can be represented by
\begin{eqnarray}
\label{eq:47}
\tilde r &=&\frac{R_0(1-e^2)}{1+e\cos \eta}, \\
\label{eq:48}
\omega_0 t&=&(1-e^2)^{3/2}\int_0^\eta \frac{\rmd \eta'}{(1+e\cos \eta')^2} \ ,
\end{eqnarray}
where $\omega_0>0$ is the corresponding Keplerian frequency, i.e. $\omega_0^2=GM/R_0^3$.

Working in the $(X,Y,Z)$ coordinate system, we introduce cylindrical coordinates $(R,\varphi , Z)$, so that Eq. (\ref{eq:41}) can be expressed in general as
\begin{eqnarray}
\label{eq:49}
&& \ddot R-R\dot \varphi^2+\frac{GMR}{(R^2+Z^2)^{3/2}}=F_R\ , \\
\label{eq:50}
&& \frac{1}{R}\frac{\rmd }{\rmd t}(R^2\dot \varphi)=F_\varphi \ ,\\
\label{eq:51}
&& \ddot Z+\frac{GMZ}{(R^2+Z^2)^{3/2}}=F_Z\ ,
\end{eqnarray}
where $F_R$ and $F_\varphi$ are given by
\beq
\label{eq:52}
F_R=F_X\cos \varphi+F_Y \sin\varphi, \quad F_\varphi=-F_X\sin \varphi+F_Y\cos \varphi \ .
\eeq

To compute the linear perturbation away from the Keplerian orbit due to relativistic effects, we evaluate
$F_R$, $F_\varphi$ and $F_Z$ along the unperturbed orbit and seek a solution of Eqs. (\ref{eq:49})--(\ref{eq:51}) of the form
\beq
\label{eq:53}
R=\tilde r (1+U), \quad \varphi=\eta +W, \quad Z=\tilde r H\ ,
\eeq
where $U$, $W$ and $H$ are only considered to linear order.
To simplify matters, it is convenient to express Eqs. (\ref{eq:49})--(\ref{eq:51}) in terms of the new independent variable $\eta$
instead of $t$, where $t(\eta)$ is given by Eq. (\ref{eq:48}). A lengthy, but straightforward calculation results in the linear perturbation equations \cite{bahram78}
\begin{eqnarray}
\label{eq:54}
&& \frac{\rmd^2 U}{\rmd \eta^2}-2\frac{\rmd W}{\rmd \eta}-3\frac{U}{1+e\cos\eta}={\mathcal A}\ , \\
\label{eq:55}
&&  \frac{\rmd }{\rmd \eta}\left(\frac{\rmd W}{\rmd \eta}+2U\right)={\mathcal B}\ ,\\
\label{eq:56}
&& \frac{\rmd^2 H}{\rmd \eta^2}+H={\mathcal C}\ ,
\end{eqnarray}
where
\beq
\label{eq:57}
({\mathcal A},{\mathcal B},{\mathcal C})=\frac{\tilde r^3}{L_N^2}(F_R,F_\varphi,F_Z)
\eeq
and $L_N^2=GMR_0(1-e^2)$. We impose boundary conditions on Eqs. (\ref{eq:54})--(\ref{eq:56}) such that
\begin{eqnarray}
\label{eq:58}
&& U=W=H=0\quad {\rm at}\quad  \eta=\eta_0\ , \\
\label{eq:59}
&& \frac{\rmd U}{\rmd \eta}=\frac{\rmd W}{\rmd \eta}=\frac{\rmd H}{\rmd \eta}=0 \quad {\rm at}\quad  \eta=\eta_0\ .
\end{eqnarray}
Thus Eqs. (\ref{eq:54})--(\ref{eq:55}) now take the form
\begin{eqnarray}
\label{eq:60}
&& \frac{\rmd^2 U}{\rmd \eta^2}+\frac{1+4e\cos\eta}{1+e\cos\eta}U={\mathcal A}(\eta)+2\int_{\eta_0}^\eta{\mathcal B}(\eta ')\rmd \eta '\ , \\
\label{eq:61}
&&  \frac{\rmd W}{\rmd \eta}+2U=\int_{\eta_0}^\eta{\mathcal B}(\eta ')\rmd \eta '\ .
\end{eqnarray}

We must now evaluate ${\mathcal A}$, ${\mathcal B}$ and ${\mathcal C}$ in order to solve the linear perturbation equations. The contribution of various source terms simply superpose; therefore, we limit our attention to the dominant relativistic terms up to linear order in ${\mathbf v}/c$. Hence
\beq
\label{eq:62}
{\mathbf F}\approx \frac{2G}{c^2}\frac{\dot {\mathbf J}\times {\mathbf x}}{r^3}-\frac{2}{c}{\mathbf v}\times {\mathbf B}\ ,
\eeq
where ${\mathbf B}$ is given by Eq. (\ref{eq:28}). It follows that with this ${\mathbf F}$, we have
\begin{eqnarray}
\label{eq:63}
{\mathcal A}&=& \frac{2G\cos i}{c^2L_N}\frac{J(t)}{\tilde r}\ , \\
\label{eq:64}
{\mathcal B}&=&\frac{2G\cos i}{c^2L_N^2}\left[ \tilde r \dot J-\frac{\omega_0R_0e\sin \eta}{\sqrt{1-e^2}}J(t)\right]\ ,\\
\label{eq:65}
{\mathcal C}&=&\frac{2G\sin i}{c^2L_N^2}\left[ -\tilde r \cos \eta \dot J+\frac{\omega_0R_0 \sin \eta}{\sqrt{1-e^2}}(2+3e\cos\eta )J(t)\right]\ ,
\end{eqnarray}
where $\dot J=J_1$, $J(t)=J_0+J_1 t$ and
\beq
\label{eq:66}
J(t)=J_0+\frac{J_1}{\omega_0}(\eta-2e\sin \eta) +O(e^2).
\eeq
To simplify the solution of Eqs. (\ref{eq:54})--(\ref{eq:56}), the source terms can be written as expansions in powers of the eccentricity as well as
\beq
\label{eq:67}
U=U_0+eU_1 +O(e^2)
\eeq
and similarly for $W$ and $H$ \cite{bahram78}. The boundary conditions would then apply term by term in such expansions. It can be shown that
\begin{eqnarray}
\label{eq:68}
\fl\quad
{\mathcal A}&=& \frac{2\cos i}{Mc^2}
\left\{ J_0\omega_0 +J_1\eta +e[J_0\omega_0\cos\eta +J_1(\eta \cos \eta-2\sin \eta)]+O(e^2)\right\}\ , \\
\label{eq:69}
\fl\quad
{\mathcal B}&=& \frac{2\cos i}{Mc^2}
\left\{ J_1-e[J_0\omega_0 \sin \eta+J_1(\eta \sin \eta+\cos\eta)]+O(e^2) \right\}
\ ,\\
\label{eq:70}
\fl\quad
{\mathcal C}&=& \frac{2\sin i}{Mc^2}
\left\{2J_0\omega_0\sin\eta +J_1(2\eta\sin\eta -\cos\eta)+e[3J_0\omega_0\sin\eta \cos \eta\right.\nonumber \\
\fl\quad
&& \left.
+J_1(3\eta\sin \eta \cos \eta+\cos^2\eta-6\sin^2\eta)]+O(e^2)\right\}\ .
\end{eqnarray}

Let us first consider the perturbation on an initially circular orbit. The expressions for $U_0$, $W_0$ and $H_0$ contain cumulative (secular) terms as well as harmonic terms. For instance, $U_0$ is given by
\beq
\label{eq:71}
\fl
U_0= \frac{2\cos i}{Mc^2}[2J_1(\eta-\eta_0)+J_0\omega_0+J_1\eta -(J_0\omega_0+J_1\eta_0)\cos (\eta-\eta_0)-3J_1\sin(\eta-\eta_0)]\ .
\eeq
The {\it dominant} secular terms are given by
\beq
\label{eq:72}
\fl\quad
U_0\sim \frac{6\cos i}{Mc^2}J_1\eta, \quad W_0\sim -\frac{5\cos i}{Mc^2}J_1\eta^2, \quad
H_0\sim -\frac{\sin i}{Mc^2}J_1\eta^2\cos \eta\ .
\eeq
Thus for $J_1\cos i>0$, the orbit spirals outward and the azimuthal velocity, given generally by
\beq
\label{eq:73}
R\frac{\rmd \varphi}{\rmd t}=\frac{\omega_0R_0}{\sqrt{1-e^2}}(1+e\cos\eta )\left(1+U+\frac{\rmd W}{\rmd \eta} \right),
\eeq
tends to decrease, since
\beq
\label{eq:74}
U_0+\frac{\rmd W_0}{\rmd \eta}\sim -\frac{4\cos i}{Mc^2}J_1\eta \ .
\eeq
Next, we note that to first order in eccentricity the orbital perturbation can be calculated using Eqs. (\ref{eq:68})--(\ref{eq:70}). The dominant secular terms turn out to be
\beq
\label{eq:75}
\fl
U_1\sim -\frac{3\cos i}{Mc^2}J_1\eta^2 \sin \eta, \quad
W_1\sim -\frac{6\cos i}{Mc^2}J_1\eta^2 \cos \eta, \quad
H_1\sim -\frac{\sin i}{Mc^2}J_1\eta\sin 2 \eta\ .
\eeq
In principle, one can continue this procedure and determine $U$, $W$ and $H$ to all orders in eccentricity.

In summary, the motion of a free test mass in the variable gravitomagnetic field of a central source is such that if the test particle starts from rest, it tends to move in the same sense as the source for $\dot J>0$ and in the opposite sense for $\dot J<0$.
On the other hand, if the test mass is already in almost periodic motion about the source, then for $\dot J>0$, the prograde (retrograde) motion tends to slow down (speed up) and the opposite takes place for $\dot J<0$.
Our thought experiments illustrating the analogues of Faraday's law and Lenz's law in section 3 and appendix B involve test masses starting from rest. It has not been possible to provide simple thought experiments to illustrate in a similar way the behavior of test particles that are already in orbit about the central source.

\section{Discussion}

The purpose of this work has been to provide an {\it explicit} treatment of gravitational induction.  The main ingredients of our discussion  include the acceleration term $(2/c)\partial_t {\mathbf A}$ in the GEM force law --- see for instance Eq. (\ref{eq:14}) --- and our special ansatz (\ref{eq:23}) for a gravitomagnetic vector potential ${\mathbf A}$ that varies linearly with time. It has been shown that, despite the existence of certain differences in the force law between GEM and electrodynamics, it is nevertheless possible to establish a certain analogy between gravitational induction and electromagnetic induction.

The acceleration term $(2/c)\partial_t {\mathbf A}$ in the gravitational force law has been traditionally interpreted to be in essence responsible for a Machian inductive action of accelerated masses such that a test mass would accelerate in the same direction as the acceleration of neighboring masses (see pages 100--103 of \cite{einst}). However, our analysis of geodesic motion in the special nonstationary spacetime in this paper demonstrates explicitly that this Machian interpretation cannot be generally maintained in general relativity \cite{17}. Indeed, for $\dot J>0$, nearly circular prograde orbits experience azimuthal deceleration rather than acceleration.

Astronomical sources generally have variable angular momenta.
For instance, external electromagnetic breaking torques tend to slow
the rotation rates of pulsars with constant moments of inertia.
The implications of our preliminary results for such systems are beyond the scope of this work.

\section*{Acknowledgment}
C. Chicone was supported in part by the grant NSF/DMS-0604331.

\appendix

\section{Kerr metric with $a=\zeta ct$}

The special nonstationary solution of linearized Einstein's equations has a nonzero Einstein tensor that modulo its symmetry can be expressed in spherical polar coordinates as
\beq
\label{eq:A1}
G_{r\phi}=-3\frac{G\dot J \sin^2\theta }{c^4r^2}
\eeq
away from the origin of spatial coordinates.
Here ${\mathbf J}=J(t)\hat {\mathbf z}$, $\dot J=J_1$ and all the other components of the Einstein tensor vanish for $r>0$.

It is interesting to calculate the Einstein tensor for a Kerr metric obtained by changing a constant $J$ to $J(t)=J_0+J_1t$ --- or equivalently changing the constant parameter $a$ to $(J_0+J_1t)/(Mc)$ --- with $J_1\not=0$ in Eq. (\ref{eq:17}). The result is quite complicated; to simplify matters, we make a time translation  $t\, \mapsto\, t-J_0/J_1$ in the new time-dependent Kerr metric in Boyer-Lindquist coordinates. The result is metric (\ref{eq:17}) with $a$ replaced by
\beq
\label{eq:A2}
a=\zeta ct\, ,
\eeq
where $\zeta$ is a constant dimensionless parameter given by $\zeta=J_1/(Mc^2)$. If $\zeta=0$, we recover the Schwarzschild metric for which the Einstein tensor vanishes. Thus an expansion of the Einstein tensor of this Kerr metric in powers of the small parameter $\zeta$ should indicate how close this solution is to the exterior vacuum field of a stationary source. In this way, we find that Eq. (\ref{eq:17}) with $a=\zeta ct$ has an Einstein tensor such that modulo its symmetry $G_{r\theta}=0$, $G_{tt}=O(\zeta^4)$, while $G_{t\phi}$ and $G_{\theta\phi}$ are $O(\zeta^3)$. Moreover, $G_{tr}$, $G_{t\theta}$, $G_{rr}$, $G_{\theta\theta}$ and $G_{\phi\phi}$ are all $O(\zeta^2)$, but
\beq
\label{eq:A3}
G_{r\phi}=-3\frac{GM\sin^2 \theta}{c^2r(r-2GM/c^2)}\zeta +O(\zeta^2)\ .
\eeq
To linear order in Newton's gravitational constant, Eq. (\ref{eq:A3}) reduces to Eq. (\ref{eq:A1}).

\section{Induced current in a circular loop}

Imagine a circular loop of radius $r$ consisting of a perfect fluid that completely fills a very thin tube and is at rest around a source of mass $M$ and angular momentum $J$. The plane of the loop makes an angle of $i$ with respect to the equatorial plane of the source as in Figure 2.

\begin{figure}
\begin{center}
\includegraphics[scale=0.35]{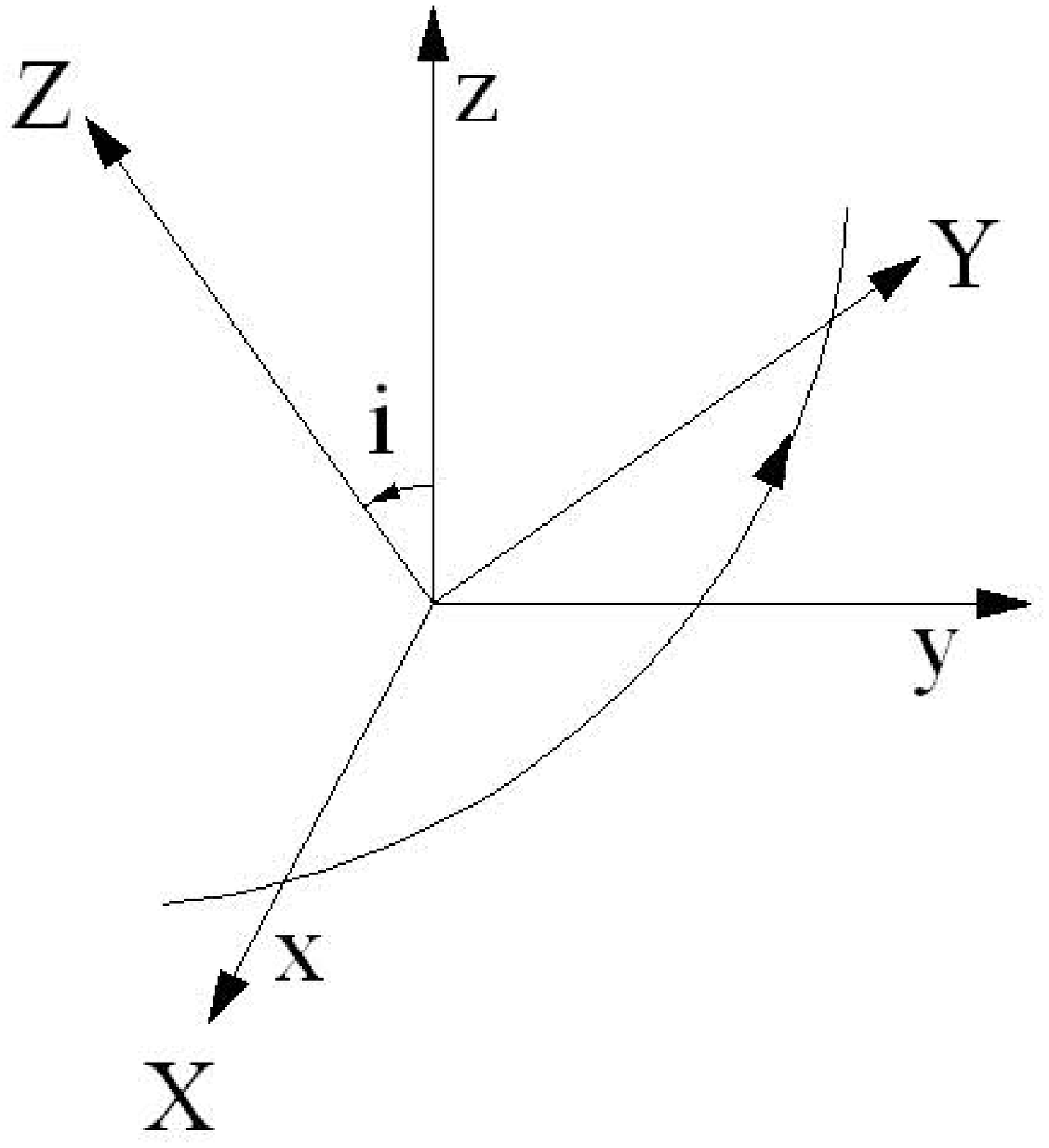}
\caption{Schematic plot of an unperturbed Keplerian orbit in the $(X,Y)$ plane. The spatial part of the background global inertial frame is given by the $(x,y,z)$ coordinate system. With respect to this, the $(X,Y,Z)$ system is rotated by a constant inclination angle $i$ about the $x$ axis.}
\end{center}
\end{figure}

We are interested in the current that is induced in the loop when $J$ is linearly dependent upon time.
The circular loop can be considered to be the boundary of a hemisphere of radius $r$ above the plane of the loop; we use the surface of the hemisphere for the calculation of the flux of the gravitomagnetic field of the source through the loop. Using the spherical polar coordinates associated with the $(X,Y,Z)$ system and
\beq
\label{eq:b1}
\hat {\mathbf z}=\cos i \, \hat {\mathbf Z}+ \sin i \, \hat {\mathbf Y}\ ,
\eeq
the flux through the loop turns out to be
\beq
\label{eq:b2}
{\mathcal F}=\frac{2\pi}{cr}GJ(t)\cos i \ .
\eeq
Similarly, the gmf can be simply calculated from Eq. (\ref{eq:34}), and we find
\beq
\label{eq:b3}
{\mathcal G}=-\frac{2G}{c^2}\oint \frac{\dot {\mathbf J}\times {\mathbf x}}{r^3}\cdot \rmd {\pmb \ell}=-\frac{4\pi}{c^2r}
G\dot J \cos i .
\eeq
Thus Eq. (\ref{eq:36}) is satisfied by the results given in Eqs. (\ref{eq:b2}) and (\ref{eq:b3}).

It is clear from Eq. (\ref{eq:29}) that the fluid particles start from rest and move in the prograde sense for $\dot J>0$ due to the presence of the acceleration term $(2/c)\partial_t {\mathbf A}$. The corresponding induced current will be flowing in the retrograde sense and is such that its negative gravitomagnetic flux through the hemisphere  opposes the increasingly positive flux of the source. The induced current is proportional to $\cos i$, so that it is maximum for the loop in the equatorial plane and vanishes for a polar loop.

It is important to remark that it is not possible to use the plane of the loop that passes through the source for the calculation of the flux. To see this, imagine instead of the surface of the hemisphere of radius $r$, the \lq\lq sombrero" surface given by the annular region of inner radius $r_S$ and outer radius $r$ together with an upper hemisphere of radius $r_S$ that just avoids the source. The flux through each part of this surface is ${\mathcal F}_A$ and ${\mathcal F}_S$, respectively, where
\beq
\label{eq:b4}
{\mathcal F}_A=\frac{2\pi}{c}\left(\frac{1}{r}-\frac{1}{r_S} \right)GJ(t)\cos i , \qquad {\mathcal F}_S= \frac{2\pi}{cr_S}GJ(t)\cos i \ .
\eeq
Note that for the {\it effective} source of the nonstationary solution, Eq. (\ref{eq:24}), $r_S\to 0$. Nevertheless, the net flux is ${\mathcal F}={\mathcal F}_A+{\mathcal F}_S$ given by Eq. (\ref{eq:b2}).


\section*{References}

\begin{thebibliography}{00}

\bibitem{ciufowhee}
Ciufolini I and  Wheeler J A 1995
{\it Gravitation and Inertia} (Princeton: Princeton University Press)

\bibitem{harris}
Harris E G 1991
{\it Am. J. Phys.} {\bf 59} 421

\bibitem{costaetal}
Costa L F and  Herdeiro C A R
{\it Preprint} gr-qc/0612140
\bibitem{bracavtho}
Braginsky V B Caves C M and Thorne K S 1977
{\it Phys. Rev. D} {\bf 15} 2047

\bibitem{mashh2}
Mashhoon B Gronwald F and  Lichtenegger H I M  2001
Lect. Notes Phys. {\bf 562}  83

\bibitem{mashh1}
Mashhoon B 2001
{\it Gravitoelectromagnetism}, in: {\it Reference Frames and Gravitomagnetism},
edited by J.-F. Pascual-S\'anchez, L. Floria, A. San Miguel and F. Vicente (Singapore: World Scientific), pp. 121--132;
2007 {\it Gravitoelectromagnetism: A Brief Review}, in: {\it The Measurement of Gravitomagnetism: A Challenging Enterprise}, edited by L. Iorio (New York: NOVA Science), ch. 3; 2005 {\it Int. J. Mod. Phys. D} {\bf 14}, 2025

\bibitem{jfps}
Pascual-S\'anchez J-F 2000
{\it Nuovo Cimento B}  {\bf 115}  725

\bibitem{mlrtart1}
Ruggiero M L and Tartaglia A 2002
{\it Nuovo Cimento B} {\bf 117}  743

%
\bibitem{iorio}
Iorio L 2002
{\it Int. J. Mod. Phys. D} {\bf 11} 781

\bibitem{mlrtart2}
Tartaglia A and Ruggiero M L 2004
{\it Eur. J. Phys.} {\bf 25} 203

\bibitem{mas93}
Mashhoon B 1993
{\it Phys. Lett. A}  {\bf 173} 347

\bibitem{co-mas}
Cohen J M and Mashhoon B 1993
{\it Phys. Lett. A} {\bf 181} 353;
Mashhoon B Iorio L and Lichtenegger H 2001  {\it Phys. Lett. A} {\bf 292} 49;
Mashhoon B  Gronwald F and Theiss D S 1999 {\it Ann. Phys. (Leipzig)} {\bf 8} 135;
Lichtenegger H Iorio L and Mashhoon B 2006  {\it Ann. Phys. (Leipzig)} {\bf 15} 868

\bibitem{bijamas}
Bini D Jantzen R T and Mashhoon B 2001
{\it Class. Quantum Grav.} {\bf 18} 653;
2002 {\it Class. Quantum Grav.} {\bf 19}  17

\bibitem{einst}
Einstein A 1950
{\it The Meaning of Relativity} (Princeton: Princeton University Press)

\bibitem{mastobepub}
Mashhoon B 2008
{\it Class. Quantum Grav.} {\bf 25} 085014

\bibitem{bahram78}
Mashhoon B 1978
{\it Astrophys. J.} {\bf 223} 285


\bibitem{17}
Lichtenegger H and Mashhoon B 2007 {\it Mach's Principle}, in: {\it The Measurement of Gravitomagnetism:
A Challenging Enterprise}, edited by L. Iorio (New York: NOVA Science), ch. 2

\end{thebibliography}
\end{document}